# Chess-board acoustic crystals with momentum-space nonsymmorphic symmetries

Yanqiu Wang[1,2,4]†, Chen Zhang[1,3,4]†, Zhiyi Chen[1,3,4], Bin Liang[1,2,4]*, Yuxin Zhao[1,3,4]*, Jianchun Cheng[1,2,4]*

**Affiliations:**

[1] Department of Physics, Nanjing University, Nanjing 210093, China.

[2] Key Laboratory of Modern Acoustics, MOE, Institute of Acoustics, Nanjing University, Nanjing 210093, China.

[3] National Laboratory of Solid State Microstructures, Nanjing University, Nanjing 210093, China.

[4] Collaborative Innovation Center of Advanced Microstructures, Nanjing University, Nanjing 210093, China.

†These authors contributed equally to this work.

*Corresponding author. Email: liangbin@nju.edu.cn

  Corresponding author. Email: zhaoyx@nju.edu.cn

  Corresponding author. Email: jccheng@nju.edu.cn

**Abstract:** Spatial symmetries appearing in both real and momentum space are of fundamental significance to crystals. However, in the conventional framework, every space group in real space, either symmorphic or nonsymmorphic, corresponds to a symmorphic dual in momentum space. Our experiment breaks the framework by showing that in a 2D acoustic crystal with chess-board pattern of $\pi$ and $0$ fluxes, mirror reflections are manifested nonsymmorphically as glide reflections in momentum space. These momentum-space nonsymmorphic symmetries stem from projective, rather than ordinary, representations of the real-space symmetries due to the peculiar flux pattern. Moreover, our experiment demonstrates that the glide reflection can reduce the topological type of the Brillouin zone from the torus to the Klein bottle, resulting in novel topological phases with new topological invariants. Since crystalline topologies are based on momentum-space symmetries, our work paves the way for utilizing engineerable gauge fluxes over artificial crystals to extend the current topological classifications into the broader regime of momentum-space nonsymmorphic symmetries.

**One-Sentence Summary:**

We show new symmetries and insulators, revealing unexplored topological phases in $k$-space nonsymmorphic symmetry classes.

Crystals are elegant forms of matter with unique spatial symmetries. Each crystal possesses a space group, which is a collection of spatial transformations such as translations, rotations, and reflections that keep the crystal unchanged (*1*). Space groups are categorized as either symmorphic or nonsymmorphic. A nonsymmorphic group contains at least one nonsymmorphic symmetry that involves a fractional lattice translation combined with rotation or reflection, whereas symmorphic groups do not involve fractional lattice translations. Scientists typically use the Fourier transform to analyze the energy spectra of crystals in momentum space. However, according to standard theory, every real-space group - whether symmorphic or nonsymmorphic - is transformed into a symmorphic group on the reciprocal lattice. Therefore, no fractional reciprocal-lattice translation is involved in any symmetry in momentum space.

In three dimensions, there are 230 space groups, of which only 73 are symmorphic. Real-space nonsymmorphic groups have been instrumental in topological phases because they can cause band crossings and lead to interesting topological phases like hourglass topological insulators (*2,3*). Therefore, proposing momentum-space nonsymmorphic groups would significantly expand the current framework of crystal symmetry and create vast opportunities to extend the current classifications of crystalline topological phases. Therefore, proposing momentum-space nonsymmorphic groups would significantly expand the current framework of crystal symmetry and create vast opportunities to extend the current classifications of crystalline topological phases.

This postulation can be achieved using projective real-space crystal symmetry, which goes beyond the standard theory of crystal symmetry. In the projective representation of the symmetry group, the combination of two symmetry operators is modified by an additional phase factor, i.e., $\rho(g_1)\rho(g_2) = e^{i\phi(g_1,g_2)}\rho(g_1 g_2)$, providing an opportunity to realize nonsymmorphic symmetries in momentum space (*4*). Projective symmetry has been incorporated to broaden the paradigm of symmetry-protected topological phases (*5,6*), which is currently a highly active field in both condensed-matter and classical-wave systems (*7-16*). Despite this, the richness of nonsymmorphic symmetries in momentum space remains unexplored, and there exists a critical gap as no experimental study of nonsymmorphic symmetry in momentum space has ever been conducted.

Artificial crystals have recently demonstrated the ability to engineer gauge fluxes 0 and $\pi$ (*17-22*), thanks to their manipulation of effective hopping phases (*23-25*) and multiple degrees of freedom (*26-28*). This engineering of gauge fluxes can projectively modify the algebras of crystal symmetries. In light of this, to realize momentum-space glide reflections, we propose a flux pattern similar to a chessboard, as shown in Fig. 1A. In this pattern, black and white squares host fluxes $\pi$ and 0, respectively. We interweave this chessboard model via introducing negative intercell coupling among the acoustic crystal in a staggered fashion, as illustrated in Fig.1B. With proper dimerization patterns, we can realize two momentum-space nonsymmorphic groups: *i*) momentum-space group *Pg*, which has only the glide reflection $\mathcal{G}_x$ along the $k_x$ direction (see Fig.1C), and *ii*) momentum-space group *Pgg*, which has both glide reflections $\mathcal{G}_x$ and $\mathcal{G}_y$ along $k_x$ and $k_y$ directions, respectively (see Fig.1D).

The two momentum-space nonsymmorphic groups possess the potential to unveil new physics. Notably, a single glide reflection $\mathcal{G}_x$ can further reduce the Brillouin torus to a Klein bottle, as depicted in Fig. 1E. Such a change in the topological type of momentum-space unit can significantly alter the topological classification from the ground up. We present compelling experimental evidence for observing a novel Klein-bottle topological insulator on acoustic crystals, which are nearly impossible to observe in the natural materials (*29-33*). Additionally, two glide reflections can lead to four non-center high-symmetry points in the bulk of the Brillouin zone (BZ) that remain invariant under $\mathcal{G}_x \mathcal{G}_y$, as highlighted in Fig.1F. The positions of the high-symmetry points serve as a distinct signature for the nonsymmorphic *Pgg*, since for all symmorphic momentum-space groups, high symmetry points locate either at the center or on the boundary of the BZ. Furthermore, we have determined the topological classification for *Pgg*, and formulate the topological invariant in terms of parities of the numbers of valence states with negative eigenvalues of $\mathcal{G}_x\mathcal{G}_y$ at these high-symmetry positions.

The momentum-space groups *Pg* and *Pgg* give rise to extraordinary edge states in topological insulators. For *Pg* with $\mathcal{G}_x$, the *y*-edges host in-gap states, while the *x*-edges remain gapped (*4*). In contrast, conventional $M_x$-symmetric topological insulators host in-gap states on the symmetry-preserving *x*-edges. In the case of *Pgg*, all four edges exhibit in-gap states, as illustrated in Fig.1G, since the *x* and *y* directions now possess equal status. Nevertheless, the momentum-space glide reflections nontrivially relate edge states: Translating the edge spectrum on one *y*-edge (*x*-edge) by $G_b/2$ ($G_a/2$) leads to exactly the same edge spectrum on the other *y*-edge (*x*-edge), as depicted in the bottom half of Fig.1G.

The experimental demonstration of momentum-space glide reflections and the Klein-bottle topological insulators on acoustic crystals opens the avenue for exploring momentum-space nonsymmorphic symmetries and their topologically protected phases through the use of engineerable gauge fluxes of artificial crystals.

**Momentum-space nonsymmorphic symmetries and topological invariants**

We now explain how the chess-board flux pattern leads to momentum-space glide reflections. It is significant to observe that the peculiar flux pattern leads to the projective algebraic relations between lattice translations $L_j$ and the mirror reflections $M_i$:

$$\{M_x, L_y\} = \{M_y, L_x\} = 0. \tag{1}$$

To see this, we note that both $M_x L_y^{-1} M_x L_y r_i$ and $M_y L_x^{-1} M_y L_x r_i$ enclose odd number of $\pi$ fluxes for an arbitrary lattice site $r_i$, and therefore $M_x L_y M_x L_y^{-1} = M_y L_x M_y L_x^{-1} = -1$. After the Fourier transform, $L_j = e^{i\mathbf{k}\cdot\mathbf{a}_j}$ where $\mathbf{a}_{x,y}$ are the lattice vectors. Hence, $M_x e^{ik_y a_y} M_x = -e^{ik_y a_y} = e^{i(k_y+G_y/2)a_y}$, and similarly $M_y e^{ik_x a_x} M_y = e^{i(k_x+G_x/2)a_y}$, with the reciprocal lattice constants $G_j = 2\pi/a_j$. Hereafter, let us set $a_j = 1$ for simplicity. Given the fractional reciprocal lattice constants in the exponents, one may immediately realize that $M_j$ act as glide reflections $\mathcal{G}_j$ on momentum space:

$$\begin{aligned}\mathcal{G}_x: (k_x, k_y) &\mapsto (-k_x, k_y + \pi),\\ \mathcal{G}_y: (k_x, k_y) &\mapsto (k_x + \pi, -k_y).\end{aligned} \tag{2}$$

Thus, we have shown that the projective algebraic relations (1) give rise to momentum-space glide reflection symmetries.

A feature of a glide reflection is that there is no fixed point under its transformation. Such a symmetry with no fixed point is referred to as being free, and can be further applied to reduce the Brillouin zone. As illustrated in Fig.1E, the fundamental domain is shrunk to half of the Brillouin zone, and $\mathcal{G}_x$ leads to anti-periodic boundary conditions for the two $k_x$ boundaries. Hence, the Brillouin torus is reduced to the Klein bottle. We recall that the topological classification over the Brillouin torus is $\mathbb{Z}$ as characterized by the Chern number. As a non-orientable manifold, the Chern number over the Klein bottle must vanish. Instead, the topological classification is drastically changed to $\mathbb{Z}_2$ over the Klein bottle, and the topological invariant can be formulated as

$$\nu = \frac{1}{2\pi i}\int_{\tau_{1/2}} d^2k\,\mathcal{F} + \frac{1}{\pi i}\oint dk_x\,\mathcal{A}_x(k_x, \tfrac{\pi}{2}) \mod 2, \tag{3}$$

where $\tau_{1/2}$ is the half of the Brillouin zone illustrated in Fig.1E, and the integration in the second term is over the upper edge of $\tau_{1/2}$. Here, $\mathcal{A}_i = \sum_\alpha \langle \psi_\alpha(k)|\partial_{k_i}|\psi_\alpha(k)\rangle$ is the Berry connection of the valence states $|\psi_\alpha(k)\rangle$, and $\mathcal{F} = \partial_{k_x}\mathcal{A}_y - \partial_{k_y}\mathcal{A}_x$ is the Berry curvature.

With both momentum-space glide reflections $\mathcal{G}_x$ and $\mathcal{G}_y$, the topological classification is still $\mathbb{Z}_2$. But, with two glide reflections, we can simplify the formula of the topological invariant. For this purpose, let us look into the combination $\mathcal{G}_x\mathcal{G}_y$, which transforms the momentum space as

$$\mathcal{G}_x\mathcal{G}_y: (k_x, k_y) \mapsto (-k_x - \pi, -k_y + \pi). \tag{4}$$

Although both $\mathcal{G}_x$ and $\mathcal{G}_y$ act freely on momentum space, it is significant to observe that the action of $\mathcal{G}_x\mathcal{G}_y$ is not free, and there are four fixed points $(\pm\pi/2, \pm\pi/2)$. These noncentral high-symmetry points in the bulk of the BZ are quite remarkable, since in ordinary theory high-symmetry momenta appear either at the center or on the boundary of the BZ. Let us label the fixed points by $K_a$ and $K_a'$ with $a = 1,2$ as illustrated in Fig.1F. $K_a$ and $K_a'$ are related by either $\mathcal{G}_x$ or $\mathcal{G}_y$.

Since $(\mathcal{G}_x\mathcal{G}_y)^2 = 1$, $\mathcal{G}_x\mathcal{G}_y$ is a twofold symmetry with eigenvalues $\pm 1$. Accordingly, for a given Hamiltonian with an energy gap, valence states at each high symmetry point can be decomposed into eigenspaces with eigenvalues $\pm 1$. Let $N_-^{K_a}$ and $N_-^{K_a'}$ denote the number of negative eigenstates at $K_a$ and $K_a'$, respectively. Since $K_a$ and $K_a'$ are symmetry related, they have the same number of negative eigenstates, i.e., $N_-^{K_a} = N_-^{K_a'}$ with $a = 1, 2$. Consequently, we can consider the parity $\nu_a$ of each number, namely

$$\nu_a = (-1)^{N_-^{K_a}}. \tag{5}$$

Then, the aforementioned $\mathbb{Z}_2$ topological invariant, namely Eq. 3, over the Klein bottle is in fact a product of $\nu_1$ and $\nu_2$, and therefore can acquire a simple formula:

$$(-1)^\nu = (-1)^{N_-^{K_1} + N_-^{K_2}}. \tag{6}$$

The proof of the simplified formula can be found in the supplementary Materials. Nevertheless, we note that the case of $\nu_1 = \nu_2$ corresponds to an atomic insulator. That is, the inclusion of $\mathcal{G}_y$ does not change the $\mathbb{Z}_2$ topological classification.

**Model with momentum-space nonsymmorphic symmetries**

We proceed to demonstrate our theory by the minimal model on the chess-board lattice. The momentum-space Hamiltonian is given by

$$\mathcal{H}(\mathbf{k}) = \begin{bmatrix} \varepsilon & q_{1,-}^x(k_x)^* & 0 & q_+^y(k_y) \\ q_{1,-}^x(k_x) & \varepsilon & q_-^y(k_y) & 0 \\ 0 & q_-^y(k_y)^* & -\varepsilon & q_{2,+}^x(k_x) \\ q_+^y(k_y)^* & 0 & q_{2,+}^x(k_x)^* & -\varepsilon \end{bmatrix}. \tag{7}$$

Here, $q_{a,\pm}^x(k_x) = t_{a1}^x \pm t_{a2}^x e^{ik_x}$ with $a = 1,2$, $q_\pm^y(k_y) = t_1^y \pm t_2^y e^{ik_y}$, and $\varepsilon$ is the on-site energy. The unit cell and the hopping amplitudes $\pm t_{ab}^x$, $\pm t_a^y$ are marked in Fig.1B, with $t_{ab}^x > 0$ and $t_a^y > 0$. One can immediately check that the gauge connections, namely the signs of the hopping amplitudes, are in accord with the chess-board flux configuration in Fig.1A. The corresponding real-space tight-binding model is given in the supplementary materials.

As seen in Fig.1C, this model automatically preserves $M_x$, and therefore it preserves the momentum-space glide reflection $\mathcal{G}_x$. Let us consider the phase transition at the 1/2-filling. When the conditions: on-site energy $\varepsilon = 0$, $t_{11}^x = t_{22}^x$ and $t_{12}^x = t_{21}^x$ are satisfied, the system is at the phase transition point if $(t_{11}^x)^2 + (t_2^y)^2 = (t_{12}^x)^2 + (t_1^y)^2$. In the vicinity of the transition point, the topological invariant $\nu$ is nontrivial if $(t_{11}^x)^2 + (t_2^y)^2 > (t_{12}^x)^2 + (t_1^y)^2$, and is trivial otherwise.

To further introduce $\mathcal{G}_y$ or $M_y$ invariance in the model, we only need to require $t_{11}^x = t_{21}^x = t_1^x$ and $t_{12}^x = t_{22}^x = t_2^x$, as illustrated in Fig.1C. We now consider the phase transition at the 1/4-filling, which occurs if $t_2^x t_2^y = t_1^x t_1^y$ with $\varepsilon = 0$. Around the transition point, the system is topologically nontrivial (trivial) if $t_2^x t_2^y > t_1^x t_1^y$ ($t_2^x t_2^y < t_1^x t_1^y$). The values of the topological invariant can be derived from Eq. 6. At the 1/4-filling, there is only one valence band, and the eigenvalues of $\mathcal{G}_x \mathcal{G}_y$ at $k = K_1$ and $K_2$ are negative and positive, respectively, if $t_2^x t_2^y > t_1^x t_1^y$, and are both positive if $t_2^x t_2^y < t_1^x t_1^y$.

**The experimental realization in the chess-board acoustic crystals**

We have experimentally verified the theoretical findings discussed above via a reconfigurable acoustic crystal. Our designed acoustic lattice structure, as photographed in Fig.2A, is carefully crafted to emulate the minimal model described by Eq. 7. The unit comprises four identical cuboid cavities (64×32×10 mm) that are connected by different square cross-sectional tubes measuring 32 mm in length. These cavities are capable to exhibit dipole-like resonance mode at the frequency $f = 2687.5$ Hz. Based on the dipole-like field distribution, we adjust the cross-section of tubes and relative position of interfaces to control the strength and relative sign of coupling between adjacent cavities, respectively. Furthermore, these narrow tubes are all designed as interchangeable components, allowing for the easy-to-assemble design of two practical samples with wholly distinct dimerization patterns.

To realize the chess-board flux configuration model, it is crucial to engineer the negative hopping amplitudes. In the left part of Fig. 2A, we illustrate that this is achieved by introducing intercellular negative couplings between cavity A and B along the $x$ direction, and between cavity B and C along the $y$ direction at the same time. In this way, π-flux is thus diagonally synthesized across these four-sites plaquettes.

We initially focus on the dimerization pattern $i$), which involves alternative dimerization along the $x$ direction and staggered dimerization along the $y$ direction. In this case, the coupling strength acting by tubes in the unit cell is arranged as $t_{11}^x = t_{22}^x = 21.4$, $t_{12}^x = t_{21}^x = 74.8$, $t_{y1} = 42.8$ and $t_{y2} = 32.1$ (Hz), respectively. Both analytical calculation and full-wave finite element method (FEM) simulation results are provided to verify our theoretical design of the target acoustic crystals auxiliary. Via simulation, we plot the bulk band structure in the first BZ of the proposed acoustic crystal in the form of iso-surface, as shown in Fig. 2B, which clearly reveals the characteristic nonsymmorphic symmetry $Pg$. Besides, Figure. 2C presents the projected band structure for the lattice along the $x$ direction computed utilizing the above-mentioned parameters, with a flat band plotted in a dashed line and the existence of topologically trivial edge states.

We now proceed to experimentally demonstrate this type of nonsymmorphic space group $Pg$ in the momentum space with coupling dimerization pattern $i$) in our fabricated acoustic crystal. The sample consists of 256 cavities arranged in 8 unit cells along both $x$ and $y$ direction, and the corresponding width of tubes is set as follows: $w_{11}^x = w_{22}^x = 4.0$, $w_{12}^x = w_{21}^x = 7.5$, $w_{y1} = 5.7$ and $w_{y2} = 4.9$ (mm). We first place a loudspeaker in the bulk cavity in the sample and scan acoustic pressure signal in each cavity. The details regarding eigenfrequency distribution in the first BZ are then extracted by applying Fourier transformation to the original acoustic field signals in the sample. Figure.2E displays an experimentally obtained contour line of eigenfrequency $f =$ 2696 Hz in the first BZ, which closely matches the predicted bulk iso-eigenfrequency contour plotted with a dashed line. It exhibits glide reflection along $k_x = 0$, which is a typical feature of $Pg$ nonsymmorphic symmetry. To further validate our findings, we move the loudspeaker to cavities located at the bottom edge along $x$ direction and detect the acoustic response in the boundary of the sample at frequency ranging from 2400 to 2900 Hz. A similar measurement is performed in the bulk region for comparison. As shown in Fig.2D, there are two main peaks in both the dark curve (bulk measurement) and light curve (edge measurement), whose valley lies within the broad bandgap in the calculated spectrum in Fig.2C. The similarity between two curves serves as solid evidence that there does not exist any in-gap state in the acoustic crystal $i$), further proving gaps in the spectrum are topologically trivial.

Under the same artificial gauge configuration, we then investigate an acoustic crystal with a different dimerization pattern $ii$). We switch to the second set of coupling tubes and carry out a series of measurements parallel with the previous experiments, where dimerization pattern $ii$) is realized through the arrangement of coupling strength in both the $x$ and $y$ direction to be staggered. In this case, the width of tubes in the unit cell is designed as $w_{11}^x = w_{21}^x = 5.1$, $w_{12}^x = w_{22}^x = 7.2$, $w_{y1} = 3.6$, and $w_{y2} = 6.2$ (mm) in order, which is visualized in the inset in Fig.3A. The change in the dimerization pattern results in the emergence of $Pgg$ nonsymmorphic symmetry in the momentum space. Figure. 3B illustrates the simulated iso-frequency distribution ranging from 2664 to 2724 Hz, where two perpendicular glide reflections against $k_x = 0$ and $k_y = 0$ are conspicuously displayed. Correspondingly, we do a 2D Fourier transform of experimentally measured pressure field and acquire a contour of the eigenfrequency $f = 2688$ Hz, as shown in Fig.3E. Due to inevitable fabrication error and practical alignment of recorded signals, our experimental result shows a slight blue shift in frequency compared to the simulated one (indicated by a dashed line).

To further investigate the emergence of topological edge states in the acoustic crystal with dimerization pattern $ii$), we conduct separate acoustic response tests in the bulk and edges along both $x$ and $y$ direction. We initially calculate the projected dispersion spectra for the $x$ direction (see Fig.3C) and $y$ direction (see Fig.3D) to predict the presence of topologically nontrivial edge states. In this parameter set, there are in-gap edge states (colored in red) existing along $x$ and $y$ edge in the first and third bandgap simultaneously, i.e. boundary localization appears at both 1/4 and 3/4-filling levels. As illustrated in Fig.3F and 3G, two peaks are observed around 2610 and 2750 Hz in the measured edge wave profiles, whose appearance is attributed to the existence of in-gap topological edge states in both boundaries. Meanwhile, two valleys are observed in Fig.3H in the bulk profile tested within a broader frequency range, and they locate in the bandgap regions (highlighted in shadow). This can be attributed to the narrowness of the second bandgap (~10 Hz) compared to the first and third bandgaps (~ 70 Hz), making it difficult to distinguish from the bulk. Therefore, only two obvious peaks are observed. It is evident that valleys of the bulk transmission curve agree quite well with the peaks in the edge transmission along both the $x$ and $y$ direction, further proving that these in-gap edge states are topologically nontrivial.

**Conclusion**

This study successfully implemented momentum-space nonsymmorphic groups $Pg$ and $Pgg$ in acoustic crystals, using a chess-board flux configuration on the crystal lattice to create projective algebras of real-space symmetries. This represents a pioneering experimental demonstration of cutting-edge mathematical concepts in practical wave systems, and opens up exciting new avenues for the design of functional devices. The results deepen our understanding of the underlying physics of topological states, and the twisted edge states discovered here allow for unprecedented flexibility and versatility in the manipulation of acoustic waves. This breakthrough has the potential to lead to a range of practical applications, such as robust high-capacity sound communications and multitasking on-chip devices.

Our research not only extends the scope of momentum-space symmetries to include nonsymmorphic symmetries, but also inspires a new direction for the development of symmetry-protected topological phases. Specifically, we propose the exploration of topological phases protected by momentum-space nonsymmorphic symmetries, which is particularly promising in artificial crystals due to their engineerable gauge fluxes, as we have evinced in our experiment. Furthermore, the universality of our proposed mechanism suggests potential extensions to other areas, such as non-Hermitian topology (*34*). We believe that our work can provide valuable inspiration for exploring different bosonic systems, such as photonic, mechanical, and elastic systems, that have not been previously explored (*35-38*). Overall, our chess-board acoustic crystals demonstrate the power of using modern mathematical concepts to advance our understanding of wave systems and to develop new technologies.

**Acknowledgments:**

**Funding:**

National Key R&D Program of China (Grant No.2022YFA1404402)

National Natural Science Foundation of China (Grants No.12161160315, No.12174181, No.11634006 and No.81127901)

the High-Performance Computing Center of Collaborative Innovation Center of Advanced Microstructures and a project funded by the Priority Academic Program Development of Jiangsu Higher Education Institutions

**Author contributions:** Y.Z., B.L. and J.C. conceived the idea and supervised the project. Y.Z., C.Z. and Z.C. did the theoretical analysis. Y.W. performed the finite-element-method simulations, designed the sample and conducted the experiments. Y.W., C.Z., Y.Z. and B.L. wrote the manuscript. All authors contributed to the discussions of results and manuscript preparation.

**Competing interests:** Authors declare that they have no competing interests.

**Data and materials availability:** The experimental data and other data that support the findings of this study are available from the corresponding authors on reasonable request.

**Supplementary Materials**

Materials and Methods
Supplementary Text
Figs. S1 to S5
References 1 ~ 42

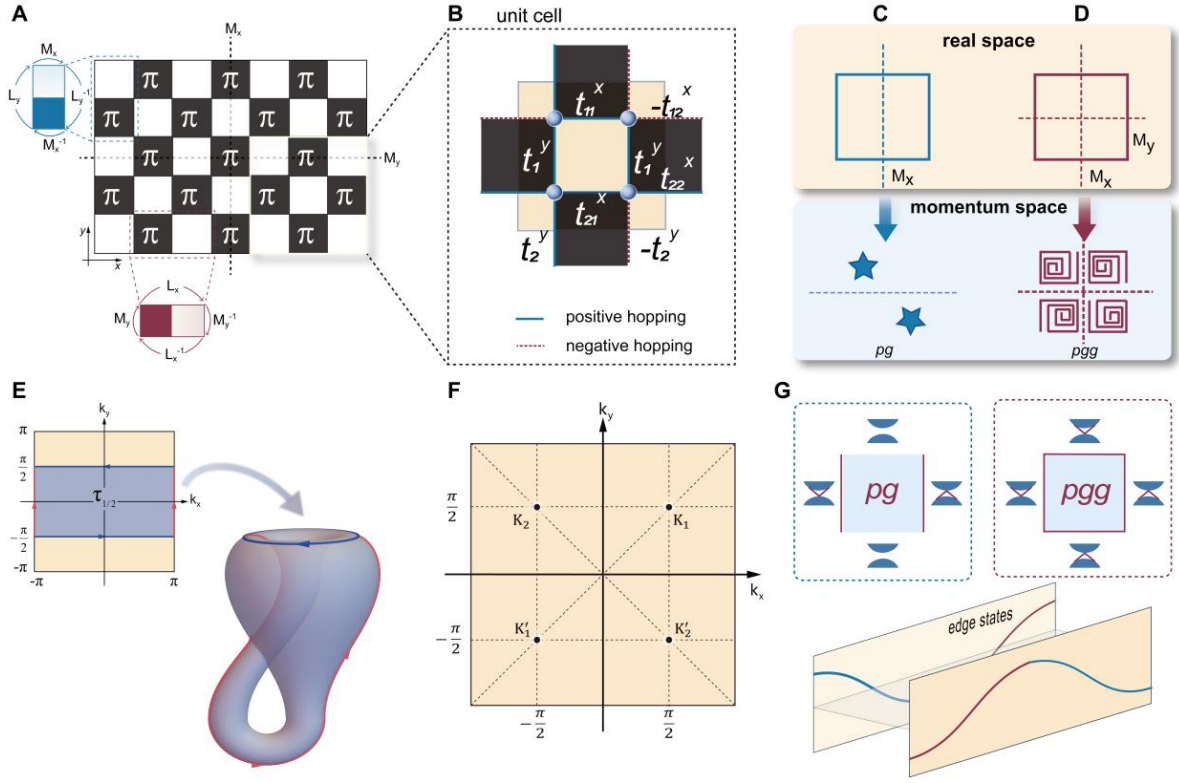

**Fig. 1. Symmetry and topology of the chess-board lattice model.** **A**, Schematic of chess-board gauge flux configuration, where shadowed plaquettes have flux $\pi$ through them, and reflection axes are marked by dashed lines. The trajectory given by $M_x L_y^{-1} M_x L_y r_j$ ($M_y L_x^{-1} M_y L_x r_j$) encircles flux $\pi \bmod 2\pi$. Here, $r_j$ is the position of an arbitrary site. **B**, The minimal model on the chess-board lattice. In accord with the gauge flux configuration, positive and negative hopping amplitudes are colored in blue and red, respectively. The unit cell is specified by the dashed blue square. **C**, The dimerization pattern for $Pg$. **D**, The dimerization pattern for $Pgg$. **E**, The Brillouin Klein bottle. The fundamental domain is chosen as $[-\pi, \pi] \times [-\pi/2, \pi/2]$ with periodic boundary conditions for two (red) $k_y$-edges and anti-periodic boundary conditions for two (blue) $k_x$-edges. The fundamental domain is topologically a Klein bottle. **F**, The momentum-space Wyckoff positions of $Pgg$, $K_1$, $K_2$, $K_1'$ and $K_2'$, invariant under $\mathcal{G}_x \mathcal{G}_y$. The four dashed lines individually preserve $\mathcal{G}_x \mathcal{G}_y$. **G**, The illustration of the edge states.

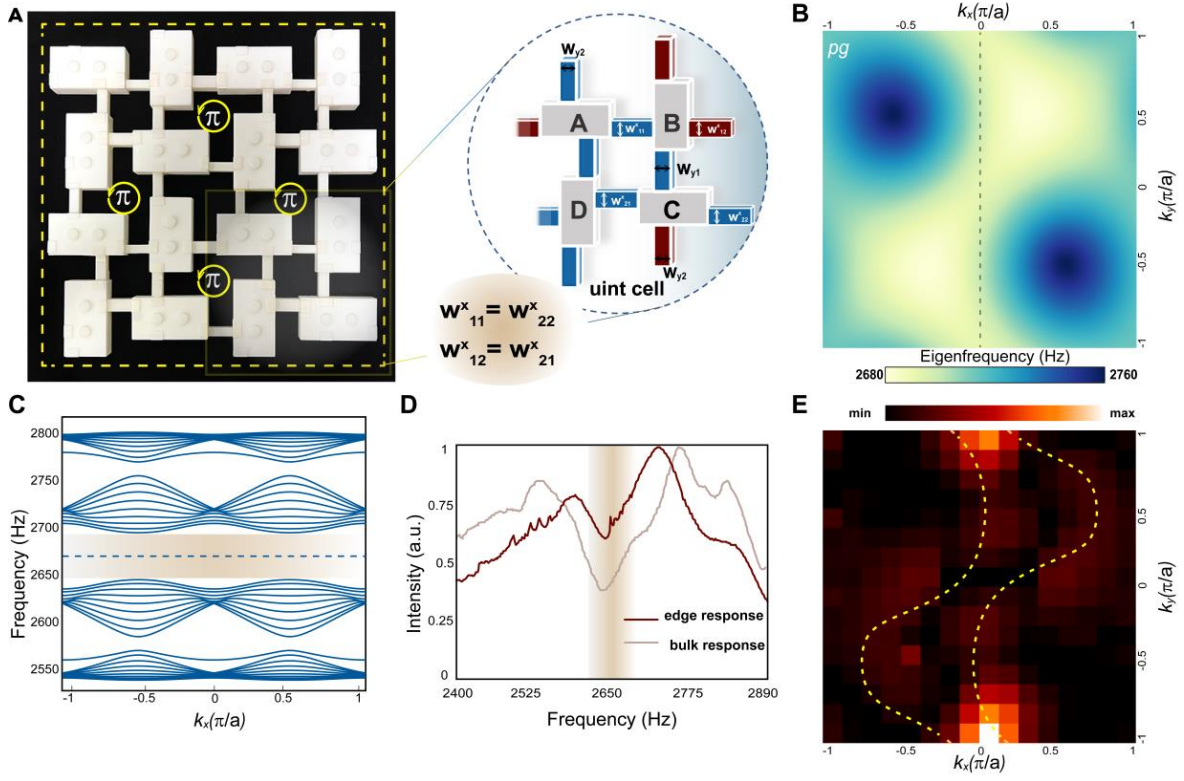

**Fig. 2. Realization of an acoustic lattice hosting nonsymmorphic symmetry *Pg* in the momentum space. A**, Photograph of the experimental realization of a chess-board-like flux configuration in the acoustic crystal. The detailed geometric parameters are marked in the enlarged sketch of unit cell, where relative positive (negative) coupling is colored in blue (red). This fabricated sample 1 conserves $M_x$ symmetry. **B**, Simulated bulk iso-eigenfrequency distribution in the *k* space with *Pg* symmetry. **C**, Analytically-calculated dispersion for the lattice with the periodic boundary condition along the *x* direction but the open boundary condition along the *y* direction. **D**, Acoustic intensity measured in the experiment when the speaker and microphone are placed at the *x* edge (light curve) and inside the bulk (dark curve). **E**, Contour line of the eigenfrequency *f* = 2696 Hz in the momentum space via 2D Fourier transform from the measured pressure field. The yellow dashed line indicates the simulated result.

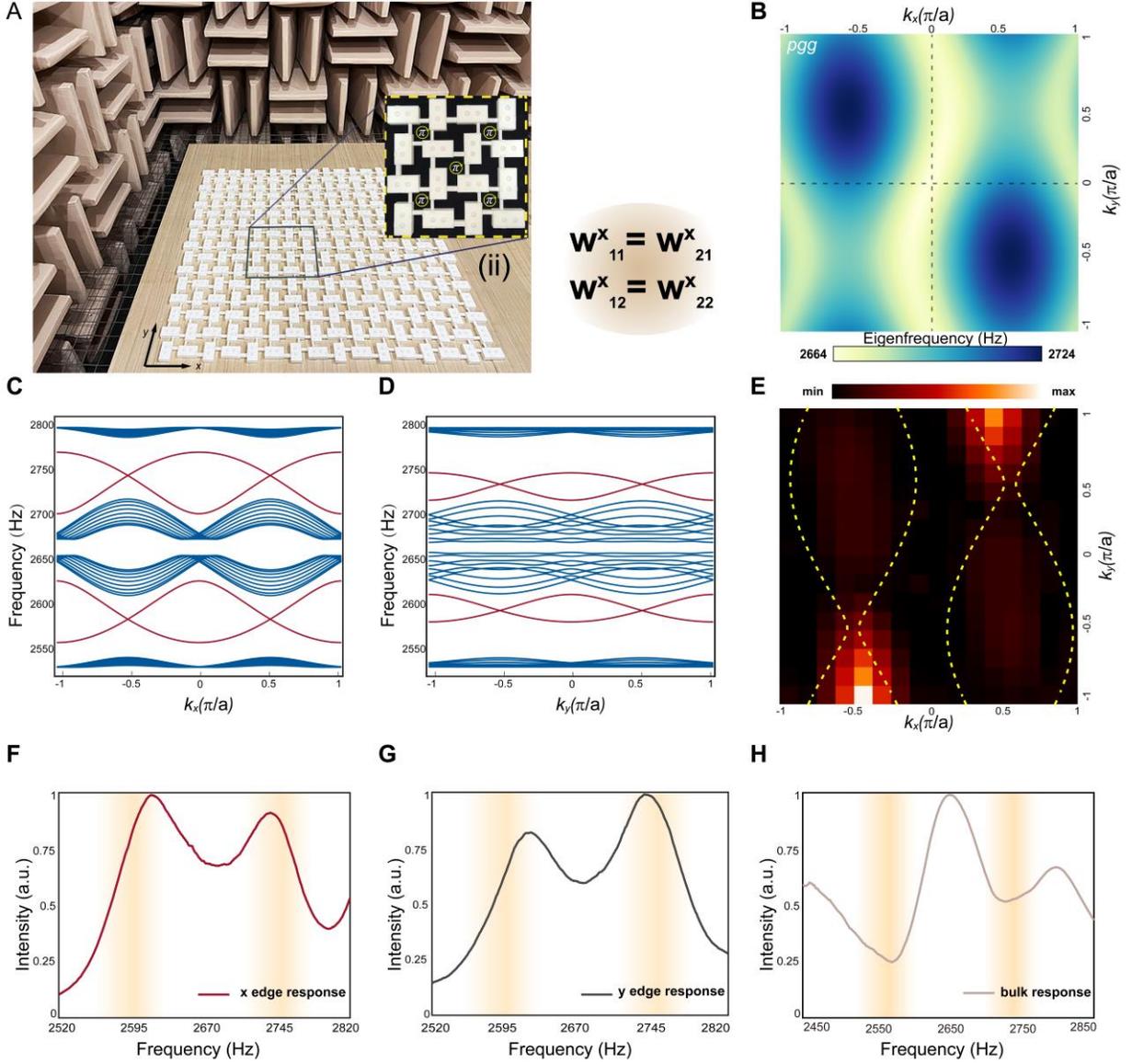

**Fig. 3. Realization of an acoustic lattice hosting nonsymmorphic symmetry *Pgg* in the momentum space.** **A**, Photograph of the fabricated acoustic crystal sample 2 shot in the anechoic chamber, with 8 unit cells in both the *x* and *y* direction. The inset shows the diagonal gauge configuration and the dimerization pattern. The fabricated sample 2 conserves $M_x$ and $M_y$ symmetries. **B**, Simulated bulk iso-eigenfrequency distribution in momentum space with *Pgg* symmetry. **C**, **D**, Analytically-calculated dispersion for the lattices supporting the edge states (red-colored bands), with the open boundary condition along the *y* (*x*) direction but the periodic boundary condition along the *x* (*y*) direction. There are two edge states colored in crimson in both directions. **E**, Measured contour line of the eigenfrequency $f = 2688$ Hz obtained from the Fourier transform of the measured pressure fields. The yellow dashed lines indicate the corresponding simulated result. **F**-**H**, Acoustic intensity measured in the experiment when the speaker and microphones are placed at two different cavities on the *x* edge, on the *y* edge, and inside the bulk, respectively.

12